\documentclass[12pt]{article}
\usepackage{graphicx}

\input{tcilatex}
\begin{document}

\bigskip \baselineskip0.8cm \textwidth16.5 cm

\begin{center}
\textbf{The estimation of (}$\mathbf{kT}_{\mathbf{C}}\mathbf{(p)/J}$\textbf{%
, }$\mathbf{p}$\textbf{) phase diagram for two-dimensional\ site-diluted
Ising model using a microcanonical algorithm}

{\small B\"{u}lent Kutlu and Ali Emre Gen\c{c}}

{\small Gazi \"{U}niversitesi, Fen Fak\"{u}ltesi, Fizik B\"{o}l\"{u}m\"{u},
06500 Teknikokullar, Ankara, Turkey}

E-mail{\small : bkutlu@gazi.edu.tr}
\end{center}

\textbf{Abstract}

{\small The site-diluted Ising model has been investigated using an improved
microcanonical algorithm from Creutz Cellular Automaton. For a
microcanonical algorithm, the basic problem is to estimate the correct
temperatures using average values of the kinetic energy in the simulations
of site-diluted Ising model. In this study, the average kinetic energy has
been re-described with an expression dependent on dilution $x=1-p$. The
values of the temperature have been calculated using the new expression and
the critical temperatures  have been estimated from the peaks of specific
heat for each value of dilution }${\small x}${\small . The obtained phase
transition line }($kT_{\mathbf{C}}(p)/J$, $p$)\textbf{\ }{\small \ is in
good agreement with functional prediction for the site-diluted Ising model.
The simulations were carried out on a square lattice with periodic boundary
conditions.}


\textbf{1 . Introduction}

In the recent years, the Ising\ model has been applied with success in many
different physical situations such as the site-diluted ferromagnet $[1-7]$,
the pure ferromagnet$[8-10]$, microemulsion$[11-13],$ structural and
magnetic phase transition$[14-19]$. Especially, the influence of defects
(site or bond dilution) on spin systems has been reviewed by many
theoretical, numerical and experimental investigations. The systems with
defect are modelled using modified versions of the most popular models for
pure systems, such as Ising and Potts models. The site-dilution on the
physical systems has a significant effect on the critical behavior of the
phase transitions. Therefore, the site-diluted Ising model has been applied
to the problems as the determination of \ the critical temperatures for the
order-disorder phase transitions $[20-22]$. \ While two-dimensional pure
Ising model was solved exactly many years ago, the diluted Ising model has
not been solved yet for two and upper dimensions. However, it has been
investigated using different simulation and approximation techniques such as
Monte Carlo $[1-5,21]$, series expansion$[7]$, renormalization group$[23,24]$%
, transfer matrix$[6]$ and mean field$[25]$. The Hamiltonian for the
site-diluted Ising model may be written as%
\begin{equation}
H_{I}=-J\sum_{<ij>}\varepsilon _{i}\varepsilon _{j}S_{i}S_{j}
\end{equation}%
where $S_{i}=\pm 1$ is the spin at site $i$ and the $J$ parameter is the
ferromagnetic spin-spin interaction energy constant. The sum is carried out
over all nearest-neighboring (nn) spin pairs on a two-dimensional square
lattice. The $\varepsilon _{i}\ $is called the occupation coefficients of
site $i$. The value of $\varepsilon _{i}\ $is $1$, if the spin is present or 
$0$ if the spin is absent at site $i$. The values of occupation coefficients
( $\varepsilon _{i}$) are randomly distributed on the lattice. Their
configurational average takes a value in the interval of $0<p=\langle
\varepsilon _{i}\rangle <1$ . On the other hand, the dilution of system is
obtained by $x=1-p.$ The dilution $x$ is $0$ for the pure Ising model. The
aim of this work is to study the critical behavior of the site-diluted Ising
model. In particular, we would like to locate the phase transition line $%
T_{C}(p)$ and to study the thermodynamic quantities at around this line. The
functional prediction for phase transition line $T_{C}(p)$ has been given by%
\begin{equation}
T_{C}(p)=\left[ \tanh ^{-1}\left( e^{-1.45\left( p-p_{c}\right) }\right) %
\right] ^{-1}
\end{equation}%
where $p_{c}=$ 0.593 $[3,26]$. Recently, the phase transition line of the
site-diluted two-dimensional Ising model has been obtained using different
Monte Carlo algorithms$[1-5]$.

In the present paper, we have studied two-dimensional site-diluted Ising
model using a cooling algorithm improved from Creutz cellular
automaton(CCA), and we have obtained the ($kT_{\mathbf{C}}(p)/J$, $p$)%
\textbf{\ }phase diagram. The CCA\ algorithm is a microcanonical algorithm
interpolating between the conventional Monte Carlo and molecular dynamics
techniques on a cellular automaton, and it was first introduced by Creutz
for the pure spin-1/2 Ising model$\left[ 27\right] $. The improved CA
algorithms for the versions of the Ising model in two and higher dimensions
have been proven to be successful in producing the values of the static
critical exponents and the critical temperatures$[28-34]$. In these
algorithms, the temperature is not an input parameter and its value is
obtained from the average kinetic energy of system. The estimation of the
correct temperature for the site-diluted system using the microcanonical
algorithms is an important problem due to dilution. The dilutions leads to
the formation of new states for kinetic energy. Therefore, the average
kinetic energy expression should be rewritten depending on dilution ( $x$
).\ In this study, we first investigated the dependence on dilution of the
average kinetic energy. Furthermore, to estimate the phase transition line $%
T_{C}(p)$, we computed the temperature variations of the specific heat ($C/k$%
) and the internal energy ($U$) for different $x=1-p$ values on the $L\times
L$ square lattice with periodic boundary conditions. \ 

The paper is organized as follows. The details of the model are given in
Section 2, the expression of average kinetic energy\ is modified for
site-diluted systems in section 3, the data are analyzed and the results are
discussed in section 4 and a conclusion is given in Section 5.

\textbf{2 . Model}

In the improved CA algorithm for site-dilute Ising model, the four variables
are associated with each site of the lattice. The value of variables on the
each site is determined from its value and those of its nearest- neighbors
at the previous time step. The updating rule, which defines a deterministic
cellular automaton, is as follows: the first two variables on each site, are
the Ising spin $B_{i}$ and the occupation coefficient $\varepsilon _{i}$.
Their values may be $0$ or $1$. The Ising spin energy for the site-diluted
model is given by Eq.1. In Eq.1 $S_{i}=B_{i}-1=\mp 1$. The third variable is
for the momentum variable conjugate to the spin ( the demon ). The kinetic
energy associated with the demon, $H_{K}$, is an integer, which equals to
the change in the Ising spin energy for any spin flip and its value lies in
the interval ($0,$ $m$). The upper limit of the interval, $m$, is equal to $%
16J$. The total energy 
\begin{equation}
H=H_{I}+H_{K}
\end{equation}%
is conserved.

The fourth variable provides a checkerboard style updating, and so it allows
the simulation of the Ising model on a cellular automaton. The black sites
of the checkerboard are updated and then their color is changed into white;
white sites are changed into black without being updated. The updating rules
for the spin and the momentum variables are as follows: For a site to be
updated, its spin is flipped and the change in the Ising spin energy ,$%
dH_{I} $, is calculated. If this change in energy is transferable to or from
the momentum variable associated with this site, such that the total energy $%
H$ is conserved, then this change is done and the momentum is appropriately
changed. Otherwise, the spin and the momentum are not changed.

For a given total energy, the temperature of the pure Ising model is
obtained from the average value of kinetic energy, which is given by$[27]$%
:\bigskip 
\begin{equation}
\langle H_{k}\rangle =\frac{\sum_{n=0}^{m}ne^{-nJ/kT}}{%
\sum_{n=0}^{m}e^{-nJ/kT}}
\end{equation}%
where $n$ is equal to the possible kinetic energy value of a site on the
spin system. The values of $n$ \ can be zero and multiples of four for the
pure Ising model. The expectation value in Eq. 4 is an average over the
lattice and the number of time steps. Because of the third variable, the
algorithm requires two time steps to give every spin of the lattice a chance
to change. Thus, in comparison to ordinary Monte Carlo simulations, two
steps correspond to one full sweep over the system variables.

The cooling algorithm is divided into two basic parts, the initialization
procedure and the taking of measurements$[28]$. In the initialization
procedure, firstly, all the spins in the lattice sites take the
ferromagnetic ordered structure ($\uparrow \uparrow $) and the occupation
coefficients ( $\varepsilon _{i}$) are randomly distributed on the lattice
corresponding to value of dilution $x$. The kinetic energy per site is equal
to the maximum change in the Ising spin energy for the any spin flip using
the third variables. This configuration is run during the 10.000 cellular
automaton time steps. In the next step, the last configuration in the
disordered structure has chosen as a starting configuration for the cooling
run. Rather than resetting the starting configuration at each energy, it is
convenient to use the final configuration at a given energy as the starting
point for the next. During the cooling cycle, energy has been subtracted
from the spin system through the third variables ($H_{k}$) after the 200.000
cellular automaton steps.

\textbf{3. The modification of the average kinetic energy expression for
site-dilute systems }

As is known, the temperature of system is obtained using average kinetic
energy for a microcanonical algorithms. For the pure Ising model, the values
of the kinetic energy ($n$) can be zero and multiples of four. But, its
values can be zero or multiples of two dependent on $x$ for site-diluted
Ising model. Therefore, to estimate the temperature of system from the
average value of the kinetic energy is a basic problem on the site-diluted
Ising model simulations with a microcanonical algorithm. At the same time,
the possible values of $n$ at each sites of the lattice vary depending on
the values of dilution $x$. If the least one of the nearest neighbor sites
of a site is empty,\ the values of $n$ can be zero or multiples of two for
this site. Otherwise, $n$ takes the values zero or multiples of four. The
dilution causes the new kinetic energy states with different frequency in
the site-diluted system compared to the the pure Ising model. Therefore, the
average value of the kinetic energy can not be defined accurately using
Eq.(4) and the system temperature can not be estimated for the site-dilute
Ising model. As a result of these, the average kinetic energy expression
should be a function of dilution $x$. As is well known, the kinetic energy
in the pure Ising model obeys the Boltzmann's law for any given temperature.
The probability $P(n)$, that $H_{K}$ has the energy $n$, is%
\begin{equation}
P(n)=\frac{\text{ }e^{-nJ/kT}}{Z}\text{.}
\end{equation}%
For the site-diluted Ising model, the expressions of the partition function
should have a different multiplier ($\rho (x)$) dependent on dilution $x$
for each kinetic energy states as given below,%
\begin{equation}
Z%
{\acute{}}%
=\sum_{z=0}^{l}\rho _{z}(x)e^{-2zJ/kT}\text{.}
\end{equation}%
In this case, the expressions of the probability and the average kinetic
energy can be re-written as a function of $x$. 
\begin{equation}
P%
{\acute{}}%
(n)=\frac{\rho _{z}(x)e^{-2zJ/kT}}{Z%
{\acute{}}%
}
\end{equation}%
\begin{equation}
\langle H_{K}\rangle =\frac{\sum_{z=0}^{l}\rho _{z}(x)\text{ }2z\text{ }%
e^{-2zJ/kT}}{\sum_{z=0}^{l}\rho _{z}(x)\text{ }e^{-2zJ/kT}}
\end{equation}%
where $z=0,1,2,...,l$ and $l$ is equal to $8$ .

Depending on the change in the Ising spin energy for a spin flip, the values
of kinetic energy can be equal to the values of $n=0,$ $4,$ $8,12,$ $16$ for
the pure system($x=0$) and the values of $n=0,$ $2,$ $...,$ $14,$ $16$ for
the site-diluted system ($x=0.5$) at any site. Here, the $2$, $6$, $10$ and $%
14$ values {}{}of kinetic energy occurs because of site-dilutions. For any
value of $x$, the $\rho _{z}(x)$ can be expressed as%
\begin{equation}
\rho _{z}(x)=(1-x)^{1-k}\text{ }x^{k}
\end{equation}%
where $k=$ $mod(z,2)$, $\rho _{z}(x)$ is equal to $1$ for even $z$ and $0$
for odd $z$ at $x\longrightarrow 0$ for the pure Ising model. However, its
values is equal to ($1-x$) for even $z$ and $x$ for odd $z$ in the
site-diluted Ising model.

\textbf{3. Results and discussion}

The thermodynamic quantities of site-diluted Ising model are calculated
using cooling algorithm. The values of the quantities are averages over the
lattice and over the number of time steps (200.000) with discard of the
first 20.000 time steps during which the cellular automaton develops. The
simulations have been performed 10 times for initial configurations with the
different total energy at each values of $x=1-p=0,$ $0.05,$ $0.10,$ $0.15,$ $%
0.20,$ $0.25$ and $0.30$ on the finite lattices with linear dimensions $%
L=60, $ $80,$ $100$ and $120$.

To determine the correct temperature value corresponding to the average
kinetic energy of system, we firstly calculated the probabilities $%
P(n)^{\ast }$ for each $n$ value from simulations on the finite size lattice
with linear dimension $L=120$. The probability of each $n$ value for kinetic
energy is calculated by

\begin{equation}
P(n)^{\ast }=N_{n}/N_{CCAS}\text{,}
\end{equation}
where $N_{n}$ is the number of \ appearance for $n$ value on kinetic energy
during the simulation and $N_{CCAS}$ is the total number of Creutz cellular
automaton steps. In order to determine the appropriate expression for the
average kinetic energy, $P(n)^{\ast }$ has been compared with the $P(n)$ and 
$P%
{\acute{}}%
(n)$ probabilities in the Eq.(5) and the Eq.(7). The variation of the
probabilities are illustrated for three different temperature values ( $%
T<T_{C}$, $T\simeq T_{C}$ and $T>T_{C}$ ) in Fig.1. It can be seen from
Fig.1 that the simulation results are in good agreement with values of the $P%
{\acute{}}%
(n)$ probability in Eq.(7) for $T<T_{C}$, $T\simeq T_{C}$ and $T>T_{C}$ . At
the same time, the calculated values of the probabilities are also in good
agreement with $P%
{\acute{}}%
(n)$ in the interval $0\leq x\leq 0.4$ for the all $\langle H_{K}\rangle $
values. This case shows that the temperature of the site-diluted system can
be estimated with the expression dependent on $x$ in Eq.(8) for average
kinetic energy.

The temperature dependence of the internal energy ($U$) and the specific
heat ($C/k$) at various values of dilution $x$ are shown in Fig.(2) and
Fig.(3) for single realizations of the site-dilution on a finite lattice
with linear dimension $L=120$. The values of the specific heat ($C/k$) and
the internal energy ($U$) \ calculated by taking the average over the time
step and lattice using the following formulas.

\begin{equation}
U=(-J\sum_{<ij>}\varepsilon _{i}\varepsilon _{j}S_{i}S_{j\text{ }})/2L^{2}
\end{equation}

\begin{equation}
C/k=L^{2}(\langle U^{2}\rangle -\langle U\rangle ^{2})/(kT)^{2}
\end{equation}%
In these figures, the values of temperatures are estimated using Eq.(8) for
the average value of the kinetic energy. Also, the simulations have been
performed 10 times with various initial configurations for the each values
of the linear dimension $L$ and of the dilution $x$. As seen in Fig.(2),
Ising energy displays a continuous behavior for all dilutions and the
functional behaviors arise in region of low temperature with increasing
dilution. In Fig.(3), the specific heat $C/k$ exhibits a strong peak around
the critical temperature for $x=0$. The peak height decreases rapidly while
the dilution increases. The functional behavior of specific heat corresponds
to second-order phase transition. At $x\geq 0.10$ region, the specific heat
shows a shoulder at temperatures above $T_{C}$ as seen in Monte Carlo
simulations$[4,5]$.

The finite-size lattice critical temperatures are also obtained from the
specific heat maxima $T_{c}^{C}(L,x)$ and the infinite lattice critical
temperatures $T_{c}^{C}(\infty ,x)$ are estimated by analyzing the data
within the framework of the finite size scaling theory. The critical
temperature values are average of the estimated critical temperature values
for the 10 different simulations. An overall error of about $3\%$ is
estimated for the values of critical temperatures. According to the finite
size scaling theory$[32]$, the infinite lattice critical temperature $%
T_{c}(\infty ,x)$ is given by,%
\begin{equation}
T_{c}(\infty ,x)=T_{c}(L,x)+aL^{-1/\nu },
\end{equation}%
where $\nu =1$ is the exponent associated with the divergence of correlation
length in infinite lattice for $x=0[30]$. The values of the finite lattice
critical temperature against $1/L$ are illustrated for various values of
dilution $x$ in Fig.4. For the all dilution values, the straight lines which
fit these data give infinite lattice critical temperature as $%
1/L\longrightarrow 0$. The critical temperature values for the finite
lattices in Fig.4(a), (b) and (c) are estimated using $n=0,2,...,16$ in
Eq.(8), $n=0,2,...,16$ and $n=0,4,...,16$ in Eq.(4) for the average value of
the kinetic energy, respectively. The estimated infinite lattice critical
temperatures obtained using Eq.(4) and Eq.(8) and the functional prediction
for phase transition line $T_{C}(p)$ are shown in Fig.5. It can be seen in
Fig.5 that the functional prediction is in good agreement with the estimated
infinite lattice critical temperatures obtained using the expression of the
average kinetic energy dependent on $x$ for the phase transition line $%
T_{C}(p=1-x)$. However, the critical temperature values estimated using $n$
values for pure Ising system are quite compatible in the interval $0\leq
x\leq 0.05$ with the predicted phase transition line. But this compatibility
disappears for the increased values of dilution ( $x\longrightarrow x_{C}$
). On the other hand, the estimated critical temperatures using $n=0,$ $2,$ $%
...,$ $14,$ $16$ values in Eq.(4) close to the predicted transition line
only for high values ( $x\longrightarrow x_{C}$ ) of dilution. The
simulations results showed that one can calculate the true temperature of
diluted system using Eq.(8) for average kinetic energy.

\textbf{4. Conclusion}

The site-dilute Ising model has been simulated using the CA cooling
algorithm for the square lattice with the ferromagnetic interactions. The
simulations show that the dilution leads to the formation of the new states
for kinetic energy in microcanonical algorithm of Ising model. Therefore,
the expression for the expected value of the kinetic energy should have a
different multiplier ($\rho (x)$) depending on dilution $x$ for each kinetic
energy state on the site-diluted Ising model. The correct temperature values
have been obtained by the expression depend on dilution ($x$). The infinite
lattice critical temperatures $T_{c}^{C}(\infty ,x)$ are estimated by
analyzing the finite-size lattice critical temperatures obtained from the
specific heat maxima within the framework of the finite size scaling theory.
For phase transition line $T_{C}($ $p=1-x)$, the estimated infinite lattice
critical temperatures using expression dependent on $x$ of the average
kinetic energy are in good agreement with the functional prediction. As a
result of calculations of probability $P(n)$, the suggested expression
(Eq.(8)) has been quite successful on the estimation of phase transition
line for site-dilute Ising model. In the simulations of the site-diluted
Ising model with a microcanonical algorithm, the system temperature can only
be estimated with the expression dependent on $x$ for the average kinetic
energy. This expression can be used for a accurate temperature account in
simulation by a microcanonical algorithm of a system containing the defect.

\textbf{Acknowledgment}

This work is supported by a grant from Gazi University (BAP:05/2010-08).

\textbf{References}

[1] I.A. Hadjiagapiou, A. Malakis and S.S. Martinos, \textit{Physica A} 
\textbf{387,} 2256 (2008).

[2] P.H.L. Martins and J.A. Plascak, \textit{Phys.Rev. E} \textbf{76} 012102
(2007).

[3] N. Schreiber and J. Adler, \textit{J.Phys.A} \textbf{38,} 7253 (2005).

[4] U.L. Fulco, F.D. Nobre, L.R. da Silva and L.S. Lucena, \textit{Physica A}%
\textbf{\ 297,} 131 (2007).

[5] W. Selke, L.N. Schur and O.A. Vasilyev, \textit{Physica A} \textbf{259,}
388 (1998).

[6] G. Mazzeo and R. K\"{u}hn, \textit{Phys.Rev. E} \textbf{60,} 3823 (1999).

[7] K. Binder, W. Kinzel and D. Stauffer, \textit{Z. Physik B} \textbf{36,}
161(1979).

[8] W. Selke and K. Binder, \textit{Ferroelectrics} \textbf{53,} 7 (1984).

[9] Q.Zhang, J. Liu and G. Wei, Phys. \textit{Stat. Sol. B} \textbf{242, }%
1093 (2005) .

[10] J. Liu, Q.Zhang, H. Yu and F. Sun, \textit{J.Magn.Magn.Mater. }\textbf{%
288,} 48 (2005).

[11] N. Jan and D. Stauffer,\textit{\ J. Phys. (France) }\textbf{49,} 623
(1988).

[12] N. Jan and D. Stauffer,\textit{\ J. Chem. Phys} \textbf{87,}6210 (1987).

[13] M. Schick and W.H. Shih, \textit{Phys. Rev. B}. \textbf{34,} 1797
(1986).

[14] W.B. Yelon, D.E.Cox, P.J. Kortman and W.B. Daniels, \textit{Phys. Rev. B%
} \textbf{9}, 4843\ (1974) .

[15] K. Binder, J. L. Lebowitz, M. K. Phani and M. H. Kalos, \textit{Acta
Metallurgica} \textbf{29}, 1655 (1981).

[16] K.E. Newman and J.D. Dow, Phys.Rev.B \textbf{27}, 7495 (1983).

[17] D. Pilay, B. Stewart, C. B. Shin and G.S. Hwang, \textit{Surface Science%
} \textbf{554, }150 (2004) .

[18] S.A.Pighin and S.A. Canas, \textit{Phys.Rev.B} \textbf{75,} 9135 (2007)

[19] G.P. Zheng, D. Gross and M. Li, \textit{Physica A}. \textbf{355,} 355
(2005).

[20] Y. Nakmura, H. Kawai and M. Nakayama, \textit{Phys.Rev.B} \textbf{55}
(1997)10549.

[21] A. Natori, M. Osanai, J. Nakamura, H.Yasunaga, \textit{Applied Surface
Science} \textbf{212}, 705(2003) .

[22] M. Osanai, H. Yasunaga and A.Natori, \textit{Surface Science} \textbf{%
493}, 319 ( 2001).

[23] A.J.E. De Souza and F.G. Brady Moreira, \textit{Europhys. Lett.} 
\textbf{17}, 491 (1992)

[24] G. G\"{u}lp\i nar and A.N. Berker, \ \textit{Phys. Rev. E} \textbf{79},
021110 (2009).

[25] H. Kaya and A.N. Berker, \ \textit{Phys. Rev. E} \textbf{62}, R1469
(2000).

[26] D.\ Stauffer and A. Aharony,\textit{\ Intoduction of Percolation Theory}
(London: Taylor and Francis) 1994.

[27] M. Creutz , \textit{Ann. Phys.} \textbf{167, }62(1986).

[28] B.Kutlu, A. \"{O}zkan, N. Sefero\u{g}lu, A. Solak and B. Binal, \textit{%
Int. J. Mod. Phys.C} \textbf{16}, 933( 2005) .

[29] B. Kutlu, S. Turan and M. Kasap,\textit{\ Int.J.Mod.Phys. C} \textbf{11,%
} 561(2000).

[30] B. Kutlu and N.Aktekin, \textit{J.Stat. Phys.}\textbf{75} 757(1994) .

[31] N.Aktekin, \textit{Physica A} \textbf{3,} 436 (1995)

[32] B.Kutlu, \textit{Int.J.Mod.Phys. C} \textbf{14}, 1305(2003) .

[33] B.Kutlu and N. Aktekin, \textit{Physica A} \textbf{215,} 370(1995) .

[34] B.Kutlu, \textit{Int. J. Mod. Phys.C} \textbf{12,} 1401( 2001) .

[35] V.Privman, \textit{Finite Size Scaling and Numerical Simulation of
Statistical systems }(World Scientific, Singapore, 1990).

\bigskip

Figure Captions:

\bigskip

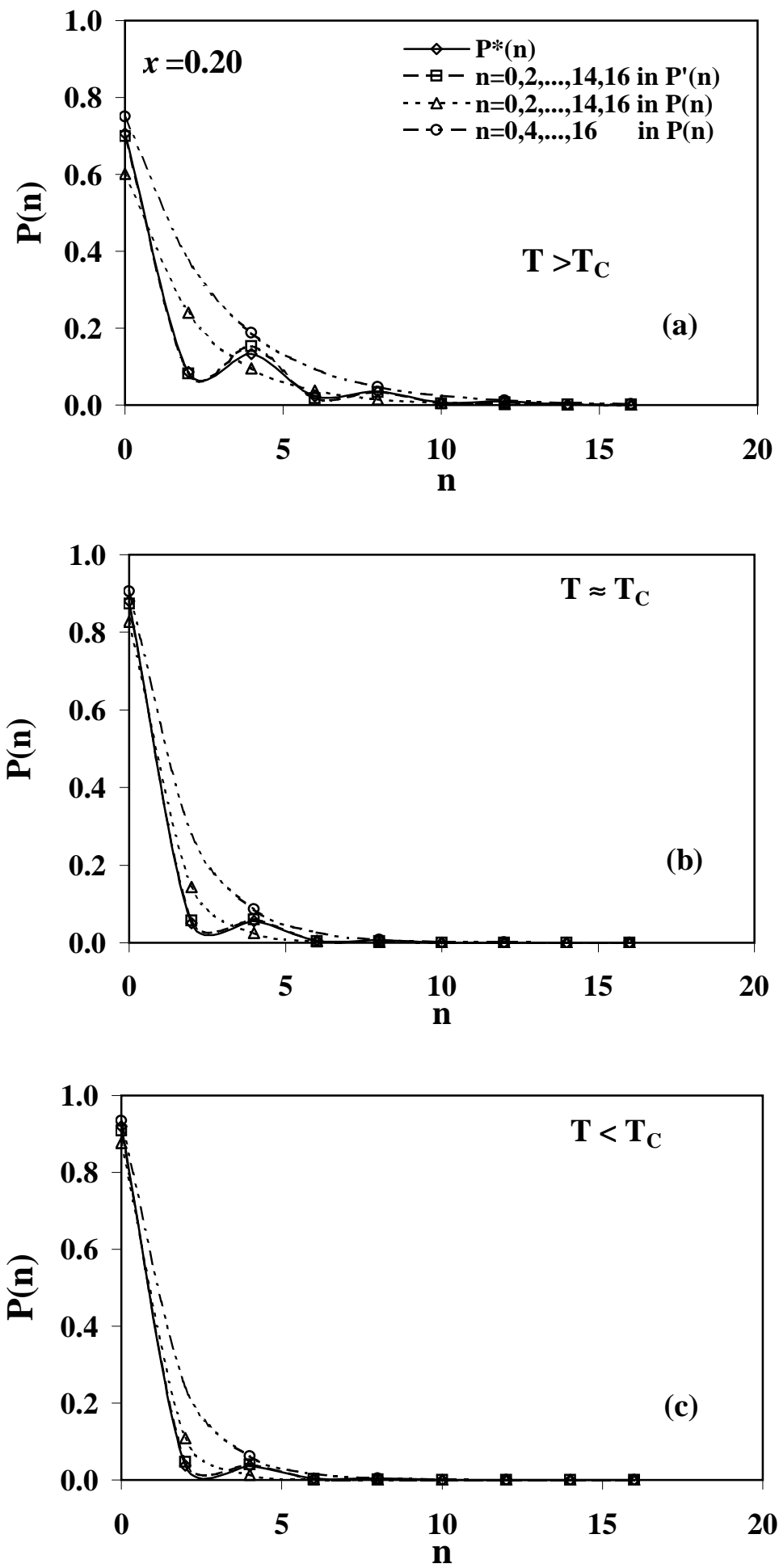
Figure 1. The $n$ dependence of the probability $P(n)$ at (a)$T>T_{C}$ (b) $%
T\simeq T_{C}$, (c) $T<T_{C}$ for $L=100$.

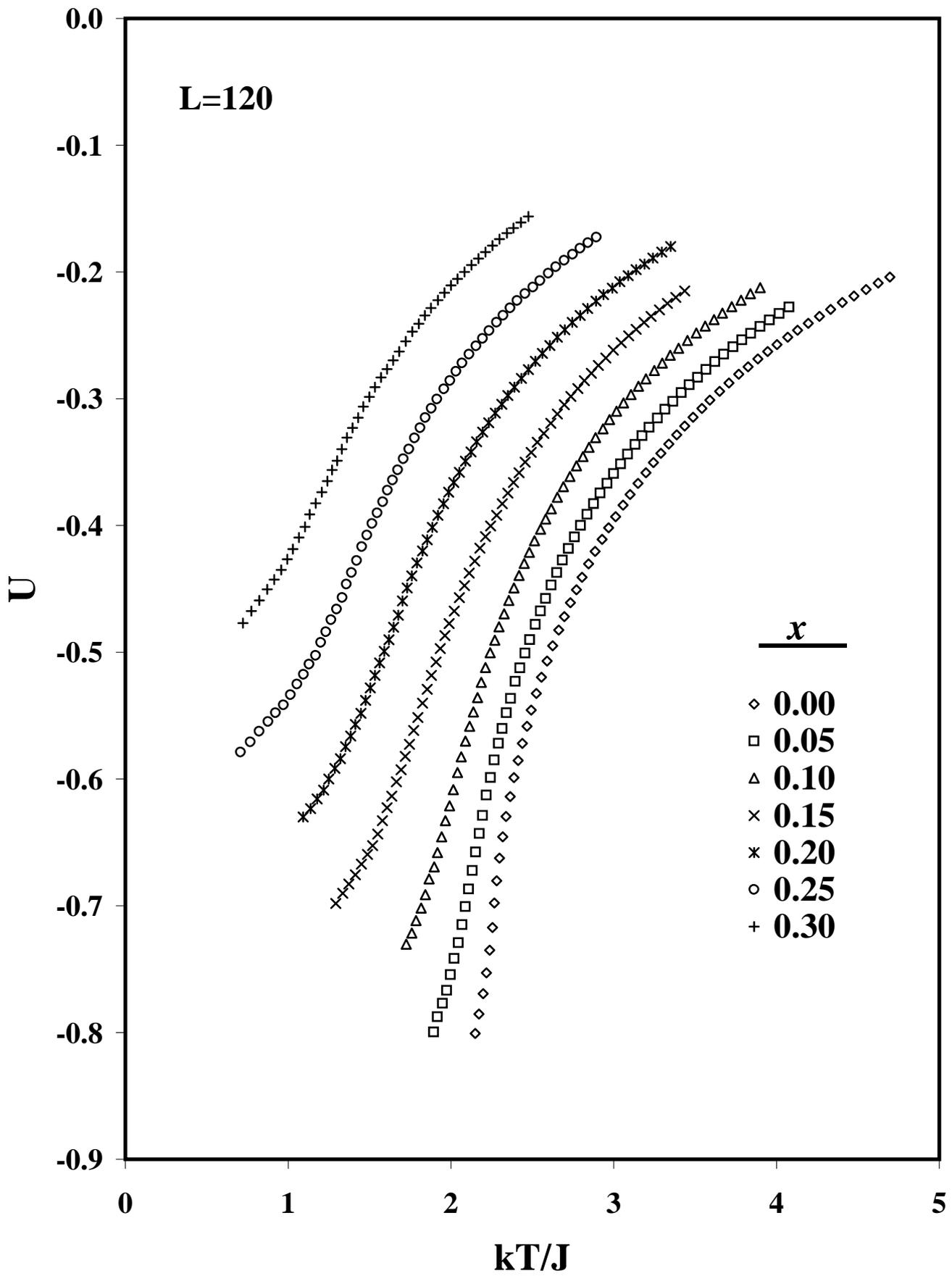
Figure 2. The temperature variation of internal energy ($H_{I}$) at the
various $x$ values for $L=120$.

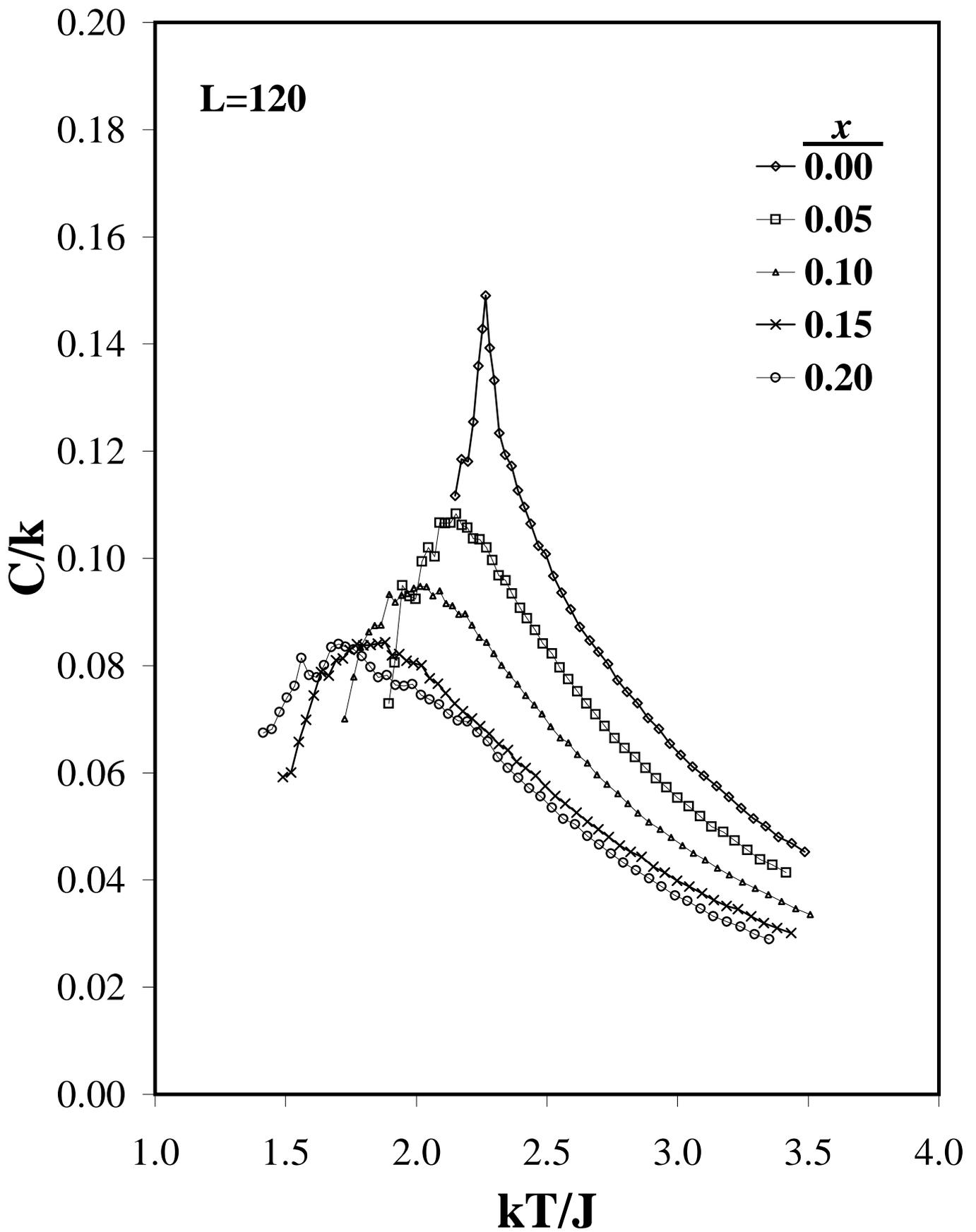
Figure 3. The temperature variation of specific heat ($C/k$) at the various $%
x$ values for $L=120$.

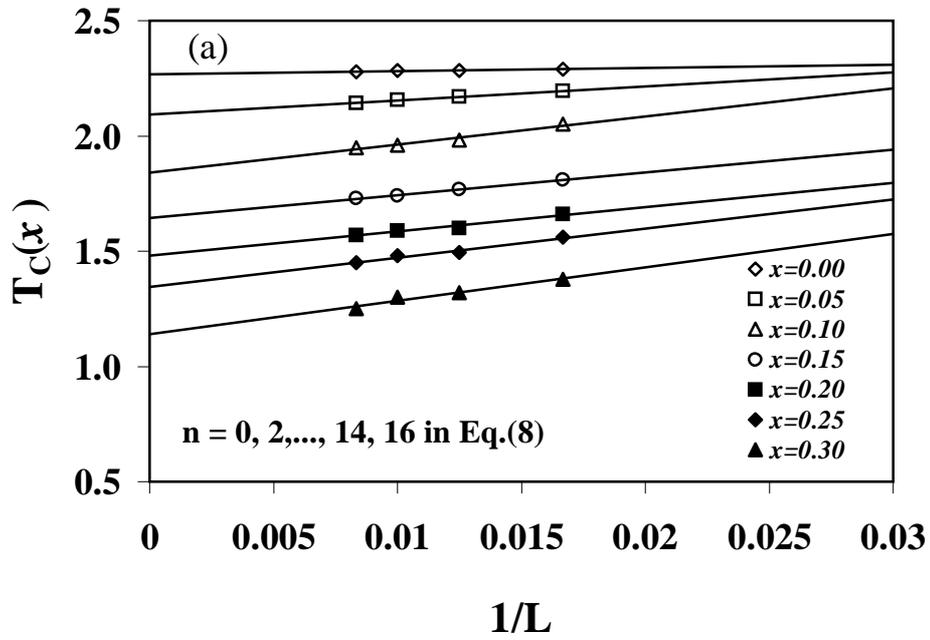
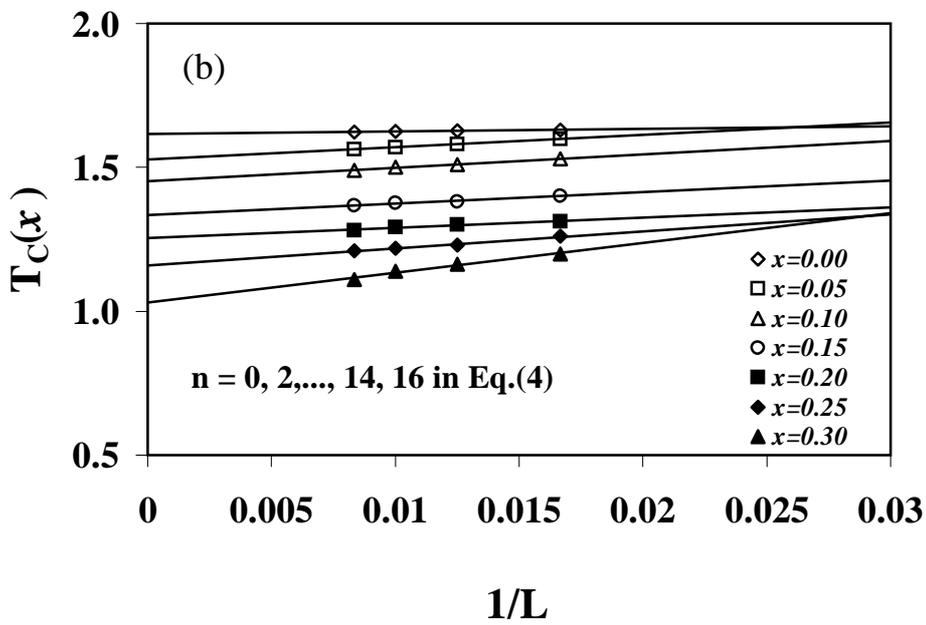
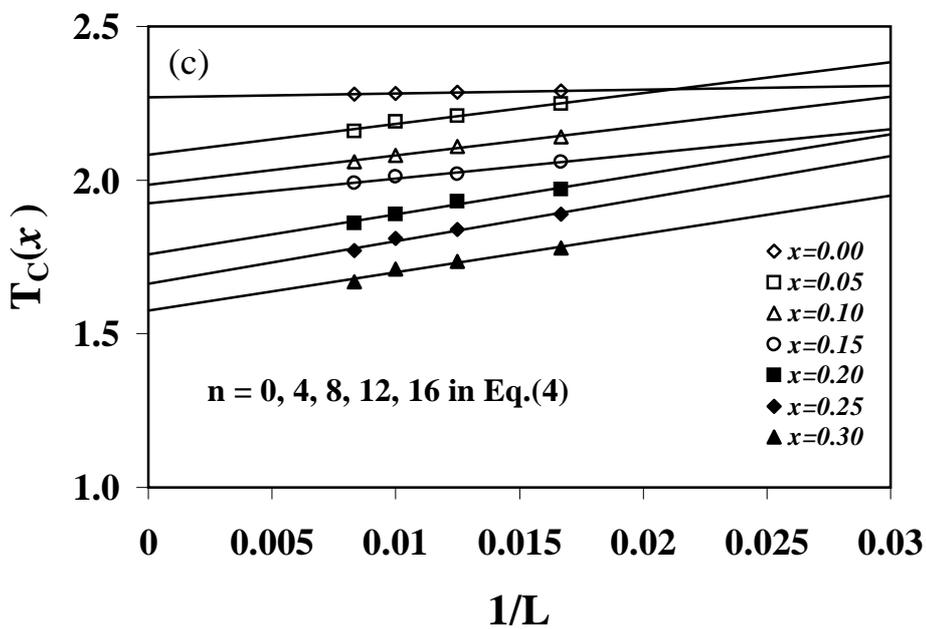
Figure 4. The plots of the finite lattice critical $T_{C}(L,x)$ temperature
against $1/L$ at various $x$ values. (a) for $n=0,2,...,14,16$ in Eq.(8),
(b) for $n=0,2,...,14,16$ in Eq.(4) and (c) for $n=0,4,8,12,16$ in Eq.(4)

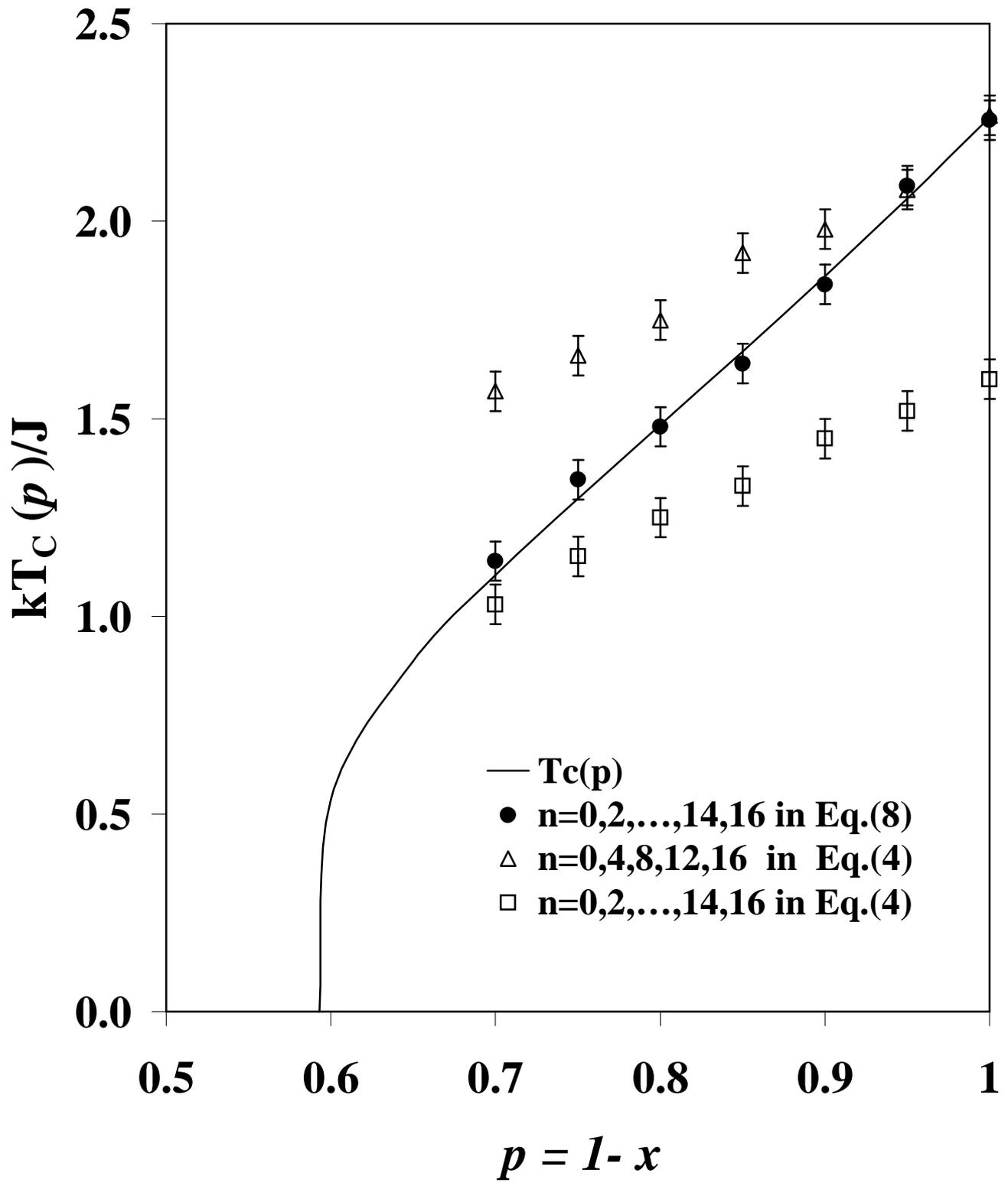
Figure 5. The phase diagram of the site diluted Ising model on square
lattice. The infinite lattice critical temperatures $T_{C}(\infty ,x)$
obtained for $n=0,2,...,14,16$ in Eq.(8), $n=0,2,...,14,16$ in Eq.(4) and $%
n=0,4,8,12,16$ in Eq.(4).

\end{document}